\documentclass[journal]{IEEEtran}
\usepackage{cite}
\usepackage{amsmath,amssymb,amsfonts}
\usepackage{algorithmic}
\usepackage{graphicx}
\usepackage{textcomp}
\usepackage{xcolor}
\usepackage{cuted}
\usepackage{epstopdf}
\usepackage{stfloats}
\usepackage{algorithm}
\usepackage{enumerate}
\usepackage{comment}

\usepackage[caption=false,font=small,labelfont=sf,textfont=sf]{subfig}

\begin{document}

\title{\fontsize{20}{26}\selectfont Low-Altitude Reflection via UAV-Mounted Rotatable IRS}
\author{Zhenwei~Jiang, Ziyuan~Zheng, Qingqing~Wu, Jing~Xu, Weiren~Zhu, Wen~Chen
\vspace{-24pt}
\thanks{Z. Jiang is with the Wireless Network Operations Center, China Unicom Ltd., Shanghai 200080, China (e-mail: jiangzhenwei1@chinaunicom.cn).}
\thanks{Z. Zheng, Q. Wu, W. Zhu, and W. Chen are with the Department of Electronic Engineering, Shanghai Jiao Tong University, 200240, China (e-mail: \{zhengziyuan2024, qingqingwu, weiren.zhu, wenchen\}@sjtu.edu.cn).}
\thanks{J. Xu is with the Shanghai Key Laboratory of Multidimensional Information Processing, East China Normal University, Shanghai 200241, China (e-mail: jxu@ce.ecnu.edu.cn)}
}

\markboth{}%
{Shell \MakeLowercase{\textit{et al.}}: Bare Demo of IEEEtran.cls for IEEE Journals}

\maketitle

\begin{abstract}
Low-altitude network is a key enabler for extending coverage and recovering connectivity in 6G systems, especially when terrestrial infrastructure is unavailable. This paper studies a uncrewed aerial vehicle (UAV)-mounted rotatable intelligent reflecting surface (IRS) as a low-altitude reflector between a blocked base station (BS) and a ground terminal (GT). Unlike the conventional isotropic-element assumption, each IRS element is modeled with a hemispherical directive radiation pattern, whose boresight can be adjusted via element rotations. We formulate a new optimization problem that jointly designs IRS phase shifts, per-element rotation vectors, and UAV placement to maximize the received signal-to-noise ratio (SNR). Leveraging the problem structure, we derive closed-form solutions for phase alignment and element rotations, showing that the optimal boresight points are along the internal angular bisector between the BS-IRS and GT-IRS directions. With these closed forms, the design reduces to a placement optimization problem over a box-constrained airspace; we solve it using an efficient projected gradient algorithm with majorization-minimization update and a global Lipschitz constant. Numerical results demonstrate substantial SNR gains from directive elements and reveal a fundamental trade-off between directional gain and path loss, yielding useful insights into low-altitude deployment of UAV-mounted IRSs.
\end{abstract}

\begin{IEEEkeywords}
Rotatable intelligent reflecting surface, uncrewed aerial vehicle, directional radiation pattern, joint rotation and placement optimization.
\end{IEEEkeywords}

\IEEEpeerreviewmaketitle

\vspace{-6pt}
\section{Introduction}
\vspace{-3pt}
Low-altitude wireless networks (LAWNs) have emerged as a promising paradigm for providing agile coverage, rapid service recovery, and capacity enhancement for 6G systems, especially in scenarios where terrestrial infrastructure is unavailable [1], [2]. Uncrewed aerial vehicle (UAV)-enabled low-altitude relaying is one of the LAWN use cases, which overcomes terrestrial blockage and provides reliable connectivity between a base station (BS) and a ground terminal (GT) [3], [4]. Compared to conventional UAV relay architectures, equipping the UAV with an intelligent reflecting surface (IRS) provides an attractive low-power option by reconfiguring wireless propagation via passive reflection, without requiring active radio-frequency chains on the UAV [5]. 
Existing UAV-mounted IRS system designs typically assume isotropic IRS elements, thereby focusing on optimizing IRS phase shifts and UAV placement to mitigate cascaded path loss [6], [7]. In practice, IRS elements often exhibit directive radiation patterns due to their electromagnetic structure, packaging, and substrate effects [8], [9]. This directionality can be exploited as an additional degree of freedom by allowing element boresight directions to be adjusted through mechanical or electronic rotations [10], [11]. This motivates us to study a UAV-mounted rotatable IRS, in which each element maintains a fixed position on the IRS aperture but can steer its boresight vector to reshape the large-scale path gains for the BS-IRS and IRS-GT links.

In this paper, we investigate a single-BS single-GT communication system with a low-altitude UAV-mounted rotatable IRS and develop a tractable joint design to maximize the received SNR. The main contributions and insights are summarized as follows: we incorporate a hemispherical cosine-based directive element pattern into the IRS reflecting model and formulate a joint optimization problem over the IRS phase shifts, element boresight rotation matrix, and UAV location for signal-to-noise (SNR) maximization; we derive the optimal rotation vector for each element in closed form under the low-altitude geometry, which admits a simple geometric form by pointing along the internal angular bisector of the two incident directions; we reduces the problem to placement-only optimization over a box region, and we propose a  closed form update rules based on the majorization-minimization (MM) and projected gradient method. Numerical results reveal a fundamental tradeoff between directional gain and path loss. Directive elements move the UAV-mounted IRS away from the location with minimum path loss in the isotropic counterpart to the location with a longer distance and larger productive path loss, but reduce the angular separation between BS and GT as seen from the IRS, thereby harvesting stronger directional gains and improving SNR. This further yields an airspace-level guideline on when directive designs are especially beneficial.

\vspace{-6pt}
\section{System Model}\label{sec:system_model}
\vspace{-3pt}
We consider a UAV-enabled IRS-aided wireless system, where a low-altitude UAV equipped with a rotatable IRS redirects communications signals from a single-antenna BS to a single-antenna GT. The direct link between BS and GT is assumed to be negligible due to severe terrestrial blockage. We adopt a global Cartesian coordinate system $\mathcal L_0 = (O,x,y,z)$, where the BS is located at the original point $O$ with coordinate $\boldsymbol{l}_{\text{B}}=\left[0,0,0\right]^T$. The coordinate of the GT is denoted by $\boldsymbol{l}_\text{T}=\left[ x_\text{T},y_\text{T},0 \right]^T$. The IRS is a uniform planar reflecting array BS with size $N$. The center location of the UAV-mounted IRS is flexible due to the mobility of the UAV in the airspace, being $\boldsymbol{l}_{\text{R}}=\left[ x_{\text{R}},y_{\text{R}},z_{\text{R}} \right]^T$, and the coordinate of each IRS element is denoted by $\boldsymbol{l}_n=\left[ x_n,y_n,z_n \right]^T,\forall n\in\mathcal{N}=\{1,\dots,N\}$. Due to airspace regulation on the placement of UAV, $\boldsymbol{l}_{\text{R}}$ is limited to a predefined area $\mathcal{A}$ by $\boldsymbol{l}_{\text{R}}\in \mathcal{A}=\{ \boldsymbol{l}\in \mathbb{R}^3|x_{\min}\le x\le x_{\max},y_{\min}\le y\le y_{\max},z_{\min}\le z\le z_{\max}\}$. 

Notably, each IRS element is implemented with a practical radiation pattern with directionality, whose boresight direction can be independently adjusted by mounting it on a rotation platform, while its physical position $\boldsymbol{l}_n$ remains fixed. The boresight of the $n$-th IRS element is characterized by a unit pointing vector $\boldsymbol{f}_n\triangleq \left[ f_{x,n},f_{y,n},f_{z,n} \right] ^T\in \mathbb{R}^{3\times 1}$ with $\lVert \boldsymbol{f}_n \rVert =1,\forall n$, where $f_{x,n}$, $f_{y,n}$, and $f_{z,n}$ denote the projections of the pointing vector on $x$-, $y$-, and $z$-axes, respectively. 
$\boldsymbol{f}_n$ can be equivalently parameterized by an elevation angle $\vartheta_n\in[0,\pi]$ between the boresight direction and the positive $z$-axis and an azimuth angle $\varphi_n\in[-\pi,\pi)$ between the projection of the boresight direction onto the $x$–$y$ plane and the positive $x$-axis as
$\boldsymbol{f}_n=\left[ \sin \vartheta _n\cos \varphi _n,\cos \vartheta _n,\sin \vartheta _n\sin \varphi _n \right]^T$. We collect all pointing vectors into a matrix $\boldsymbol{F} \triangleq [ \boldsymbol{f}_1,\dots,\boldsymbol{f}_N ] \in \mathbb{R}^{3 \times N}$, which serves as IRS rotation configuration. 
Let $\boldsymbol{e}_z\triangleq[0,0,-1]^{T}$ denote the IRS normal direction down-tilted to the ground. In general, the feasible set of boresight vectors is
\begin{align}
\mathcal{F}_n \triangleq \big\{\boldsymbol{f}_n\in\mathbb{R}^3:\|\boldsymbol{f}_n\|=1,~\arccos(\boldsymbol{f}^{T}_n\boldsymbol{e}_z)\ge \vartheta_{\max}\big\},
\label{eq:feasible_region}
\end{align}
where $\vartheta_{\max}\in(0,\pi/2]$ specifies the maximum allowable tilt angle from the array normal, and the feasible set of orientation matrices $\boldsymbol{F}$ is $\mathcal{F}$. To draw useful insights on deployment and configuration enabled by rotatable IRS elements with directional gain, we consider an ideal rotating model by setting $\vartheta_{\max}=\pi/2$, i.e., each boresight vector can freely steer within the radiating half-space, providing a performance upper bound.

The radiation pattern of each IRS element depends on the incident signal direction $\boldsymbol{r}_{\text{B},n}$ and $\boldsymbol{r}_{\text{T},n}$ from the BS and the GT, respectively, relative to its boresight $\boldsymbol{f}_n$. For simplicity of representation, we denote $\text{X}\in\{\text{B},\text{T}\}$ as either the BS or the GT, and we define the placement-aware unit direction vector
\begin{align}
\boldsymbol{r}_{\text{X},n}\triangleq\frac{\boldsymbol{l}_{\text{X}}-\boldsymbol{l}_n}{\|\boldsymbol{l}_{\text{X}}-\boldsymbol{l}_n\|}= \frac{\boldsymbol{d}_{\text{X},n}}{\lVert \boldsymbol{d}_{\text{X},n} \rVert},
\end{align}
with $\boldsymbol{d}_{\text{X},n}=\boldsymbol{l}_{\text{X}}-\boldsymbol{l}_n$ being a vector representing for a propagation path.
Then, the mismatch angle between $\boldsymbol{f}_n$ and $\boldsymbol{r}_{\text{X},n}$ can be characterized by $\psi_{n} \triangleq \arccos ( \boldsymbol{f}_n^T \boldsymbol{s}_{n}), \forall n \in \mathcal{N}$. Then, following a commonly used parametric model with a cosine-based half-space directional gain, the normalized gain pattern of each IRS element is modeled as [12]
\begin{align}
G\left( \boldsymbol{f}_{n},\boldsymbol{l}_n \right) =\begin{cases}
	G_0( \boldsymbol{f}_{n}^{T}\boldsymbol{r}_{\text{X},n} ) ^{2q},&		\boldsymbol{f}_{n}^{T}\boldsymbol{r}_{\text{X},n}\ge 0,\\
	0,&		\text{otherwise},\\
\end{cases}
\label{eq:element_gain}
\end{align}
where $q\ge 0$ is the directivity factor determining the main-lobe beamwidth, and $G_0 \triangleq 2(2q+1)$ is chosen to satisfy the power conservation law. The maximum gain $G_0$ is achieved when the incident wave is aligned with the RA boresight  $\boldsymbol{f}_{n}^{T}\boldsymbol{r}_{\text{X},n}=1$.

The propagation environment is characterized by a dominant line-of-sight (LoS) path. For the LoS path between the $n$-th IRS element and the BS or GT, the large-scale path gain is
\begin{align}
\!\!\!g_{X,n}(\boldsymbol{f}_n,\boldsymbol{l}_n)\!
\triangleq\! \frac{\beta _0 \sqrt{G_{\text{X},n}( \boldsymbol{f}_n,\boldsymbol{l}_n)}}{4\pi \lVert \boldsymbol{l}_{\text{X}}-\boldsymbol{l}_n \rVert ^2} \!=\! \frac{\beta_0 \sqrt{G_{\text{X},n}(\boldsymbol{f}_{n},\boldsymbol{l}_{n})}}{4\pi d_{\text{X},n}^2}, \!\!\! 
\label{eq:LoS_gain}
\end{align}
where $\beta_0>0$ captures the reference LoS path loss at a unit distance, the radiation pattern gain $G_{\text{X},n}( \boldsymbol{f}_n,\boldsymbol{l}_n)$ depends on the LoS direction from the BS or GT to IRS element $n$ measured relative to the boresight $\boldsymbol{f}_n$, and the propagation distance is defined by $d_{\text{X},n}=\lVert \boldsymbol{l}_{\text{X}}-\boldsymbol{l}_n \rVert$. Collectively, the overall large-scale path gain is represented by a vector
\begin{align}
\boldsymbol{g}_X\left( \boldsymbol{F} \right) =\left[ g_{X,1}^{}\left( \boldsymbol{f}_1 \right) ,\dots ,g_{X,N}^{}\left( \boldsymbol{f}_N \right) \right]^T.
\end{align}
The small-scale channel phase related to the location of each IRS element is then modeled as 
\begin{align}
h_{X,n}^{}\left( \boldsymbol{l}_n \right) \triangleq e^{j\frac{2\pi}{\lambda}d_{\text{X},n}}=e^{j\frac{2\pi}{\lambda_c}\lVert \boldsymbol{l}_{\text{X}}-\boldsymbol{l}_n \rVert},
\end{align}
where $\lambda_c$ is the wavelength, and the channel vector with normalized gain is $\boldsymbol{h}_{X}^{}\left( \boldsymbol{l}_{\text{R}} \right) =\left[ h_{X,1}^{}\left( \boldsymbol{l}_1 \right) ,\dots ,h_{X,N}^{}\left( \boldsymbol{l}_N \right) \right] ^T$. Finally, we denote the reflection coefficient of the IRS for each element as $\left( \boldsymbol{\theta } \right) =\mathrm{diag}\left( \theta _1,\dots ,\theta _N \right)$ with $ \left| \theta _n \right|=1$.

\vspace{-9pt}
\section{Problem Formulation}
\vspace{-3pt}
According to the modeling of IRS rotation, IRS reflection, and rotation- and placement-aware propagation channel, the received signal and SNR of the GT is respectively given by
\begin{align}
&\boldsymbol{u}=P_{\text{B}}\left( \mathrm{diag}\left( \boldsymbol{g}_{\text{T}} \right) \boldsymbol{h}_{\text{T}}^{} \right) ^T\mathrm{diag}\left( \boldsymbol{\theta } \right) \left( \mathrm{diag}\left( \boldsymbol{g}_{\text{B}}^{} \right) \boldsymbol{h}_{\text{B}}^{} \right) s+w
\\
&\gamma=\frac{P_{\text{B}}}{\sigma ^2}
\big| \left( \mathrm{diag}\left( \boldsymbol{g}_{\text{T}} \right) \boldsymbol{h}_{\text{T}}^{} \right) ^T\mathrm{diag}\left( \boldsymbol{\theta } \right) \left( \mathrm{diag}\left( \boldsymbol{g}_{\text{B}}^{} \right) \boldsymbol{h}_{\text{B}}^{} \right) \big|^2,
\end{align}
where $P_{\text{B}}$ is the BS transmit power, $s$ is the BS transmitted symbol and $w$ is the additive white Gaussian noise with $w\sim \mathcal{CN}(0,\sigma^2)$ and power $\sigma^2$. Now, we aim to determine the optimal configuration of UAV-mounted rotatable IRS to maximize the GT received SNR. Therefore, we formulate a joint optimization problem on the UAV location, IRS rotation, and IRS phase shifts as follows
\begin{subequations}
\begin{align}
\left( \text{P1} \right) : \underset{\boldsymbol{\theta },\boldsymbol{F},\boldsymbol{l}_{\text{R}}}{\max}&\,\,\gamma  ( \boldsymbol{\theta },\boldsymbol{F},\boldsymbol{l}_{\text{R}})
\\
\text{s.t.} \quad &  \left| \theta _n \right|=1,\forall n\in \mathcal{N},
\\
&    \lVert \boldsymbol{f}_n \rVert =1,\forall n\in \mathcal{N},
\\
&        \boldsymbol{l}_{\text{R}}\in \mathcal{A},
\end{align}
\end{subequations}
where (9b) is the unit-modulus constraint, (9c) is the rotation definition, and (9d) is the placement regulation constraint. This problem is challenging due to highly coupled variables and a non-convex, non-concave objective function, which is non-trivial and cannot be solved directly by existing methods.
To efficiently obtain the solution to (P1), we first explicitly expand the expression of the SNR (9a) as below
\begin{align}
&\gamma ( \boldsymbol{\theta },\boldsymbol{F},\boldsymbol{l}_{\text{R}}) \!=\!\frac{\beta _{0}^{2}G_{0}^{2}P_{\text{B}}}{256\pi ^4\sigma ^2}\Big|\! \sum_{n\in \mathcal{N}}{\!\frac{\big[( \boldsymbol{l}_{\text{B}}\!-\!\boldsymbol{l}_n ) ^T\!\boldsymbol{f}_n \big]_+^{q}\big[( \boldsymbol{l}_{\text{T}}\!-\!\boldsymbol{l}_n ) ^T\!\boldsymbol{f}_n \big]_+^{q}}{\lVert \boldsymbol{l}_{\text{B}}-\boldsymbol{l}_n \rVert ^{q+2}\lVert \boldsymbol{l}_{\text{T}}-\boldsymbol{l}_n \rVert ^{q+2}}}
\nonumber
\\
&\qquad \qquad \qquad\qquad\qquad \quad \times \!\theta _n e^{j\frac{2\pi}{\lambda}\left( \lVert \boldsymbol{l}_{\text{B}}-\boldsymbol{l}_n \rVert +\lVert \boldsymbol{l}_{\text{T}}-\boldsymbol{l}_n \rVert \right) } \Big|^2\!,\!\!
\end{align}
where $[\cdot]_+$ represents for $\max\{\cdot,0\}$ corresponding to the half-space directive gain (3). In the following, we derive useful insights from (10) that motivate the algorithm design . 

\vspace{-6pt}
\section{Proposed Solution}
\vspace{-3pt}
From (10), we observe that although the variables are coupled, the IRS phase shift $\boldsymbol{\theta}$ only relates to placement $\boldsymbol{l}_\text{R}$, and each phase shift $\theta_n$ can be designed separately for joint phase alignment by letting
\begin{align}
\theta^\star_n=e^{-j\frac{2\pi}{\lambda}\left( \lVert \boldsymbol{l}_{\text{B}}-\boldsymbol{l}_n \rVert +\lVert \boldsymbol{l}_{\text{T}}-\boldsymbol{l}_n \rVert \right)},\forall n\in \mathcal{N}.
\end{align}
With the optimal given $\boldsymbol{\theta}^\star$, the SNR is only related to $\boldsymbol{l}_\text{R}$ and $\boldsymbol{F}$ and written as
\begin{align}
\!\!\!\gamma ( \boldsymbol{F}\!,\boldsymbol{l}_{\text{R}}|\boldsymbol{\theta}^*) \!=\!\chi _0\bigg| \!\sum_{n\in \mathcal{N}}{\!\!\frac{\big[ ( \boldsymbol{l}_{\text{B}}\!-\!\boldsymbol{l}_n )^T \! \boldsymbol{f}_n \big]_+ ^{q}\big[ ( \boldsymbol{l}_{\text{T}}\!-\!\boldsymbol{l}_n )^T \! \boldsymbol{f}_n \big]_+^{q}}{\lVert \boldsymbol{l}_{\text{B}}-\boldsymbol{l}_n \rVert ^{q+2}\lVert \boldsymbol{l}_{\text{T}}-\boldsymbol{l}_n \rVert ^{q+2}}} \bigg|^2\!, \!\!\!
\end{align}
where $\chi_0=\frac{\beta _{0}^{2}G_{0}^{2}P_{\text{B}}}{256\pi ^4\sigma ^2}$ represents the constant coefficient in (10).

For (12), which only contains large-scale path gain, although the expression is in a form of indeterminate power, $\gamma$ w.r.t. $\boldsymbol{F}$ has a unique structure w.r.t. IRS element rotations $\boldsymbol{f}_n$, as compared to $\boldsymbol{l}_\text{R}$ with fractions. Specifically, given a fixed $\boldsymbol{l}_\text{R}$ and $\boldsymbol{\theta}^\star$, the rotation optimization subproblem is
\begin{align}
\!\!\!(\text{P2}):\underset{\|\boldsymbol{f}_n\|=1,\forall n}{\max}\,&\bigg| \sum_{n\in \mathcal{N}}{\!\frac{1}{d_{\text{B},n}^{q+2}d_{\text{T},n}^{q+2}}[ \boldsymbol{d}_{\text{B},n}^{T}\boldsymbol{f}_n] _{+}^{q}[ \boldsymbol{d}_{\text{T},n}^{T}\boldsymbol{f}_n ] _{+}^{q}} \bigg|^2. \!\!\!
\end{align}
\noindent \textbf{Remark 1:} As the IRS element adopts a half-space radiation model (3), where a direction $\boldsymbol{r}_{\text{X},n}$
is radiated only if its projection onto the boresight vector $\boldsymbol{f}_n$ is nonnegative, i.e., $\boldsymbol{f}_n^T\boldsymbol{r}_{\text{X},n}\ge 0$; otherwise the radiation gain is zero.
Assume that $\boldsymbol{r}_{\text{B},n}$ and $\boldsymbol{r}_{\text{T},n}$ lie in a common radiating half-space, namely, the angle between $\boldsymbol{r}_{\text{B},n}$ and $\boldsymbol{r}_{\text{T},n}$ is less than $180^{\circ}$, there must exist a unit boresight vector $\boldsymbol{f}_n$ such that $\boldsymbol{f}_n^T\boldsymbol{r}_{\text{B},n}>0$ and $\boldsymbol{f}_N^T\boldsymbol{r}_{\text{T},n}>0$. 
This geometry requirement is inherently satisfied in the considered low-altitude relying scenarios for ground BS and GT with UAV-mounted down-tilted reflecting IRS.

Following Remark 1, we consider that the BS and GT are located at the radiation half-space of the UAV-mounted IRS, and have Proposition 1 to obtain a closed-form optimal $\boldsymbol{F}^\star$.

\noindent \textbf{Proposition 1:}
With half-space unlimited rotation range on (1) and low-altitude BS-UAV-GT geometry on (3), the optimal solution $\boldsymbol{F}^\star$ to problem (P2) is given in a closed form by
\begin{align}
\boldsymbol{f}_{n}^{\star}&=\frac{d_{\text{T},n}\boldsymbol{d}_{\text{B},n}+d_{\text{B},n}\boldsymbol{d}_{\text{T},n}}{\lVert d_{\text{T},n}\boldsymbol{d}_{\text{B},n}+d_{\text{B},n}\boldsymbol{d}_{\text{T},n} \rVert}=\frac{\boldsymbol{r}_{\text{B},n}+\boldsymbol{r}_{\text{T},n}}{\lVert \boldsymbol{r}_{\text{B},n}+\boldsymbol{r}_{\text{T},n} \rVert},
\end{align}
with the optimal value being
\begin{align}
\gamma \left( \boldsymbol{F}^{\star}|\boldsymbol{l}_{\text{R}},\boldsymbol{\theta }^{\star} \right) =\chi _0\bigg|\sum_{n\in \mathcal{N}}{\frac{\left( \boldsymbol{r}_{B,n}^{T}\boldsymbol{r}_{\text{T},n}+1 \right) ^{q}}{2^{q}d_{\text{B},n}^{2}d_{\text{T},n}^{2}}}\bigg|^2.
\end{align}
\begin{IEEEproof}
The objective value of (13) mainly depends on 
\begin{align}
&(\boldsymbol{d}_{\text{B},n}^{T}\boldsymbol{f}_n ) ^{q}( \boldsymbol{d}_{\text{T},n}^{T}\boldsymbol{f}_n ) ^{q}=\big( \boldsymbol{f}^T\boldsymbol{D}_n\boldsymbol{f} \big) ^{q},
\\
&\boldsymbol{D}_n=\boldsymbol{D}_n^T=(\boldsymbol{d}_{\text{B},n} \boldsymbol{d}_{\text{T},n}^{T}+\boldsymbol{d}_{\text{T},n}\boldsymbol{d}_{B,n}^{T})/2.
\end{align}
As the change on the rotation $\boldsymbol{f}_n$ of the $n$-th IRS element does not effect other $N-1$ summation terms w.r.t. $\boldsymbol{f}_{n^{'}}, \forall n^{'}\ne n$ in (13), (P2) can be equivalently divided into $N$ subproblems in parallel, and the $n$-th subproblem is rewritten as follows
\begin{align}
\text{(P3)}: \underset{\| \boldsymbol{f}_n \|=1}{\max}\,\,\boldsymbol{f}_{n}^{T}\boldsymbol{D}_n\boldsymbol{f}_n,
\end{align}
where we ignore the current constant terms when fixed $\boldsymbol{l}_\text{R}$ and take the monotonicity of the power function $(\cdot)^{q}$ with a positive exponent $q$. Since $\boldsymbol{D}_n$ is symmetric, the objective in (P3) is the Rayleigh quotient. Hence, we obtain
\begin{equation}
\max_{\|\boldsymbol{f}_n\|= 1}\ \boldsymbol{f}_n^T \boldsymbol{D}_n\boldsymbol{f}_n=\lambda_{\max}(\boldsymbol{D}_n),
\end{equation}
where $\lambda_{\max}(\cdot)$ denotes the maximum eigenvalue, and any unit-norm eigenvector associated with $\lambda_{\max}(\boldsymbol{D}_n)$ is optimal.
Next, exploiting the rank-two structure of $\boldsymbol{D}_n$, we note that
\begin{align}
\boldsymbol{D}_n\boldsymbol{d}_{\text{B},n}^{}=( \boldsymbol{d}_{\text{T},n}^{T}\boldsymbol{d}_{\text{B},n}^{}\cdot \boldsymbol{d}_{\text{B},n}^{}+\lVert \boldsymbol{d}_{\text{B},n}^{} \rVert ^2\cdot \boldsymbol{d}_{\text{T},n}^{} )/2, 
\\
\boldsymbol{D}_n\boldsymbol{d}_{\text{T},n}^{}=( \lVert \boldsymbol{d}_{\text{T},n}^{} \rVert ^2\cdot \boldsymbol{d}_{\text{B},n}^{}+\boldsymbol{d}_{\text{B},n}^{T}\boldsymbol{d}_{\text{T},n}^{}\cdot \boldsymbol{d}_{\text{T},n}^{} )/2. 
\end{align}
Therefore, on the subspace $\mathrm{span}\{\boldsymbol{d}_{\text{B},n},\boldsymbol{d}_{\text{T},n}\}$, $\boldsymbol{D}_n$ admits the $2\times2$ representation $\frac{1}{2}\left[ \begin{matrix}
	\boldsymbol{d}_{\text{T},n}^{T}\boldsymbol{d}_{\text{B},n}^{}&		\lVert \boldsymbol{d}_{\text{B},n}^{} \rVert ^2\\
	\lVert \boldsymbol{d}_{\text{T},n}^{} \rVert ^2&		\boldsymbol{d}_{\text{B},n}^{T}\boldsymbol{d}_{\text{T},n}^{}\\
\end{matrix} \right] $
under the basis $\{\boldsymbol{d}_{\text{T},n},\boldsymbol{d}_{\text{B},n}\}$.
Its two eigenvalues solve
\begin{align}
&\det \left| \tfrac{1}{2}\left[ \begin{matrix}
	\boldsymbol{d}_{\text{T},n}^{T}\boldsymbol{d}_{\text{B},n}^{}&		\lVert \boldsymbol{d}_{\text{B},n}^{} \rVert ^2\\
	\lVert \boldsymbol{d}_{\text{T},n}^{} \rVert ^2&		\boldsymbol{d}_{\text{B},n}^{T}\boldsymbol{d}_{\text{T},n}^{}\\
\end{matrix} \right] -\lambda \boldsymbol{I} \right| =0, 
\\
&\quad \Longleftrightarrow ( \boldsymbol{d}_{\text{B},n}^{T}\boldsymbol{d}_{\text{T},n}^{}-2\lambda ) ^2-\lVert \boldsymbol{d}_{\text{B},n}^{} \rVert ^2\lVert \boldsymbol{d}_{\text{T},n} \rVert ^2=0,
\end{align}
which yields the maximum one being
\begin{align}
\lambda _{\max}\left( \boldsymbol{D}_n \right)=(\boldsymbol{d}_{\text{B},n}^{T}\boldsymbol{d}_{\text{T},n}^{}+\lVert \boldsymbol{d}_{\text{B},n}^{} \rVert \lVert \boldsymbol{d}_{\text{T},n}^{} \rVert)/2.
\end{align}
Substituting $\lambda _{\max}\left( \boldsymbol{D}_n \right), \forall n \in \mathcal{N}$ into (13), we arrive at (15). To obtain an eigenvector associated with $\lambda _{\max}\left( \boldsymbol{D}_n \right)$, consider
$\boldsymbol{\mu}_n\triangleq \lVert \boldsymbol{d}_{\text{T},n} \rVert \boldsymbol{d}_{\text{B},n}+\lVert \boldsymbol{d}_{\text{B},n} \rVert \boldsymbol{d}_{\text{T},n}$.
Using these relationships,
\begin{align}
\boldsymbol{D}_n&\boldsymbol{\mu}_n=\lVert \boldsymbol{d}_{\text{T},n}^{} \rVert \boldsymbol{D}_n\boldsymbol{d}_{\text{B},n}+\lVert \boldsymbol{d}_{\text{B},n}^{} \rVert \boldsymbol{D}_n\boldsymbol{d}_{\text{T},n} \nonumber
\\
=&(\boldsymbol{d}_{\text{B},n}^{T}\boldsymbol{d}_{\text{T},n}^{}\!\!+\!\| \boldsymbol{d}_{\text{B},n} \| \| \boldsymbol{d}_{\text{T},n} \|)( \|\boldsymbol{d}_{\text{T},n} \| \boldsymbol{d}_{\text{B},n}\!\!+\!\|\boldsymbol{d}_{\text{B},n} \| \boldsymbol{d}_{\text{T},n} )/2 \nonumber
\\
=&\lambda _1\left( \boldsymbol{D}_n \right) \boldsymbol{v}_n,
\end{align}
so $\boldsymbol{\mu}_n$ is an eigenvector of $\boldsymbol{D}_n$ for $\lambda _{\max}( \boldsymbol{D}_n)$.
Normalizing it by $\| \boldsymbol{f}_n \|=1$ gives (14).
\end{IEEEproof}
\noindent \textbf{Remark 2:} Proposition 1 reveals a geometric structure that for the single-BS single-GT UAV-mounted IRS communication under the cosine-based hemispherical element pattern, the unconstrained optimal boresight of each IRS element points along the \textit{internal angular bisector} of the BS-element and GT-element directions, satisfying $\angle(\boldsymbol f_n^\star,\boldsymbol{r}_{\text{B},n})
=\angle(\boldsymbol f_n^\star,\boldsymbol{r}_{\text{T},n})
=\frac{1}{2}\angle(\boldsymbol{r}_{\text{B},n},\boldsymbol{r}_{\text{T},n})$, which maximizes the product of the two directional gains
$(\boldsymbol f_n^T\boldsymbol{r}_{\text{B},n})^{q}(\boldsymbol f_n^T\boldsymbol{r}_{\text{T},n})^{q}$ under $\|\boldsymbol f_n\|=1$.

With the closed-form phase alignment $\boldsymbol{\theta}^\star$ in (11) and the closed-form element rotations $\boldsymbol{F}^\star$ in (14), the received SNR becomes a function of the IRS center location only, denoted by $\gamma(\boldsymbol{l}_{\text{R}}\mid\boldsymbol F^\star,\boldsymbol\theta^\star)$.
Accordingly, with (15), the UAV-mounted IRS placement optimization subproblem is equivalent to
\begin{align}
\!\!\!(\text{P4}): \underset{\boldsymbol{l}_{\text{R}}\in\mathcal A}{\max} \sqrt{\gamma(\boldsymbol{l}_{\text{R}}\mid\boldsymbol F^\star,\boldsymbol\theta^\star)}=\sqrt{\chi_0}\sum_{n\in \mathcal{N}}{\!\frac{( \boldsymbol{r}_{B,n}^{T}\boldsymbol{r}_{\text{T},n}\!+\!1) ^{q}}{2^{q}d_{\text{B},n}^{2}d_{\text{T},n}^{2}}}. \!\!\!
\end{align}
(P4) is non-concave w.r.t.\ $\boldsymbol{l}_{\text{R}}$ due to the undetermined power and coupled distance- and angle-dependent terms in fractions. We use a projected gradient method to obtain an efficient update under the simple box constraint $\boldsymbol{l}_\text{R} \ in \mathcal A$ and guarantee monotonicity with an MM procedure. Note that since the IRS elements form a UPA and the UAV controls only the IRS center location, shifting $\boldsymbol{l}_{\text{R}}$ translates all element coordinates by the same increment.
For a the IRS with $N=N_xN_y$ elements and inter-element spacings $\Delta_x$ and $\Delta_y$ along the $x$- and $y$-axes, respectively, the $n$-th element location satisfies
\begin{align}\label{eq:ln_from_lR}
\!\!\boldsymbol{l}_{n}\!
=\!\boldsymbol{l}_{\text{R}}\!+ \!\boldsymbol{\bar l}_n,\,\,\boldsymbol{\bar l}_n\!=\!\Big[\!\Big(\!i\!+\!\frac{N_x\!-\!1}{2}\!\Big)\Delta_x,\Big(\!j\!+\!\frac{N_y\!-\!1}{2}\!\Big)\Delta_y,0\Big]^T\!,\!\!\!
\end{align}
where $\boldsymbol{\bar l}_n$ is the fixed offset determined by the UPA geometry. Recall that
$d_{\text{X},n}=\|\boldsymbol{l}_\text{X}-\boldsymbol{l}_n\|$ and
$\boldsymbol{r}_{\text{X},n}=(\boldsymbol{l}_{\text{X}}-\boldsymbol{l}_n)/d_{\text{X},n}$ for $\text{X}\in\{\text{B},\text{T}\}$, the following derivatives w.r.t. $\boldsymbol{l}_{\text{R}}$ hold
\vspace{-2pt}
\begin{align}
&\nabla_{\boldsymbol{l}_{\text{R}}} d_{\text{X},n} = -\boldsymbol{r}_{\text{X},n}, \qquad \nabla _{\boldsymbol{l}_{\text{R}}}\boldsymbol{r}_{\text{X},n}\!=\!\frac{\boldsymbol{I}\!-\!\boldsymbol{r}_{\text{X},n}\boldsymbol{r}_{\text{X},n}^{T}}{d_{\text{X},n}},
\\
&\nabla_{\boldsymbol{l}_{\text{R}}}(\boldsymbol{r}_{\text{B},n}^T\boldsymbol{r}_{\text{T},n})\!=\!-\frac{(\boldsymbol{I}\!-\!\boldsymbol{r}_{\text{B},n}\boldsymbol{r}_{\text{B},n}^T)\boldsymbol{r}_{\text{T},n} } {d_{\text{B},n}} -\frac{(\boldsymbol{I}\!-\!\boldsymbol{r}_{\text{T},n}\boldsymbol{r}_{\text{T},n}^T)\boldsymbol{r}_{\text{B},n}}{d_{\text{T},n}}.
\end{align}
\vspace{-1pt}Then, with chain rules, the gradient of the $n$-th summand in (26) is given by (30) based on (28) and (29). We denote $v\left( \boldsymbol{l}_{\text{R}} \right) =\sqrt{\gamma \left( \boldsymbol{l}_{\text{R}}\mid \boldsymbol{F}^{\star},\boldsymbol{\theta }^{\star} \right)}$ in the following for brevity. Combining (30) for all IRS elements yields the desired
$\nabla_{\boldsymbol{l}_{\text{R}}}v(\boldsymbol{l}_{\text{R}})$ for optimization.
\begin{figure*}
\begin{align}
&\nabla _{\boldsymbol{l}_{\text{R}}}\frac{( 1+\boldsymbol{r}_{\text{B},n}^{T}\boldsymbol{r}_{\text{T},n} ) ^q}{2^qd_{\text{B},n}^{2}d_{\text{T},n}^{2}}=\frac{( 1+\boldsymbol{r}_{\text{B},n}^{T}\boldsymbol{r}_{\text{T},n}) ^{q}}{2^{q}d_{\text{B},n}^{2}d_{\text{T},n}^{2}}\Big( \frac{q}{1+\boldsymbol{r}_{\text{B},n}^{T}\boldsymbol{r}_{\text{T},n}}\big( -\frac{(\boldsymbol{I}\!-\!\boldsymbol{r}_{\text{B},n}\boldsymbol{r}_{\text{B},n}^{T}) \boldsymbol{r}_{\text{T},n}}{d_{\text{B},n}}-\frac{( \boldsymbol{I}\!-\!\boldsymbol{r}_{\text{T},n}\boldsymbol{r}_{\text{T},n}^{T}) \boldsymbol{r}_{\text{B},n}}{d_{\text{T},n}} \big) +\frac{\boldsymbol{r}_{\text{B},n}}{d_{\text{B},n}}+\frac{\boldsymbol{r}_{\text{T},n}}{d_{\text{T},n}} \Big), \!\!\!
\\
& \overline{\ \ \ \ \ \ \ \ \ \ \ \ \ \ \ \ \ \ \ \ \ \ \ \ \ \ \ \ \ \ \ \ \ \ \ \ \ \ \ \ \ \ \ \ \ \ \ \ \ \ \ \ \ \ \ \ \ \ \ \ \ \ \ \ \ \ \ \ \ \ \ \ \ \ \ \ \ \ \ \ \ \ \ \ \ \ \ \ \ \ \ \ \ \ \ \ \ \ \ \ \ \ \ \ \ \ \ \ \ \ \ \ \ \ \ \ \ \ \ \ \ \ \ \ \ \ \ \ \ \ \ \ \ \ \ \ \ \ \ \ \ \ \ \ \ \ \ } \nonumber
\end{align}
\vspace{-33pt}
\end{figure*}
To deal with the non-concavity of the objective function (26) w.r.t. $\boldsymbol{l}_\text{R}$, we resort to the MM method. At iteration $t$, based on the second-order Taylor expansion, we construct a global concave quadratic lower bound of $v$ at $\boldsymbol{l}_{\text{R}}^{(t)}$
\begin{align}
\underline{v}\big(\boldsymbol{l}_{\text{R}}\mid \boldsymbol{l}_{\text{R}}^{(t)}\big)
=&
v\big(\boldsymbol{l}_{\text{R}}^{(t)}\big)
+
\nabla_{\boldsymbol{l}_{\mathrm  R}}v\big(\boldsymbol{l}_{\text{R}}^{(t)}\big)^{T}
\big(\boldsymbol{l}_{\text{R}}-\boldsymbol{l}_{\text{R}}^{(t)}\big) \nonumber
\\
&-\frac{L}{2}\big\|\boldsymbol{l}_{\text{R}}-\boldsymbol{l}_{\text{R}}^{(t)}\big\|^{2},
\label{eq:mm_minorizer}
\end{align}
where $L>0$ is the Lipschitz constant chosen such that $\underline{v}(\boldsymbol{l}_{\text{R}}\mid \boldsymbol{l}_{\text{R}}^{(t)})\le \gamma(\boldsymbol{l}_{\text{R}})$ holds for all $\boldsymbol{l}_{\text{R}}\in\mathcal A$ [13].
Define the lower bound on the BS-IRS and GT-IRS distances over $\mathcal A$ as
\begin{align}
d_{\min}\triangleq
\min_{\boldsymbol{l}_{\text{R}}\in\mathcal A}\ \min_{n\in\mathcal N}\ \min_{\mathrm X\in\{\mathrm B,\text{T}\}}
d_{\mathrm X,n}(\boldsymbol{l}_{\text{R}}).
\label{eq:dmin_def}
\end{align}
Since $\mathcal A$ is compact and $d_{\mathrm X,n}(\boldsymbol{l}_{\text{R}})$ is continuous, $d_{\min}$ exists and is positive as long as $\mathcal A$ does not intersect the point of BS $\boldsymbol{l}_\text{B}$ and GT $\boldsymbol{l}_\text{T}$. This constant can be computed exactly offline using the point-to-box distance operation (32).

\noindent\textbf{Lemma 1:} Over $\mathcal A$, the gradient $\nabla_{\boldsymbol{l}_{\text{R}}} v(\boldsymbol{l}_{\text{R}})$ is Lipschitz continuous with a valid global constant
\begin{align}
L=\sqrt{\chi_0}N(4q^{2}+22q+20)d_{\min}^{-6}.
\label{eq:L_global}
\end{align}

\begin{IEEEproof}
Write $v(\boldsymbol{l}_{\text{R}})=\sqrt{\chi_0}\sum_{n\in\mathcal N} c_n(\boldsymbol{l}_{\text{R}})$ with
$c_n(\boldsymbol{l}_{\text{R}})\triangleq a_n(\boldsymbol{l}_{\text{R}})b_n(\boldsymbol{l}_{\text{R}})$,
where
$a_n(\boldsymbol{l}_{\text{R}})\triangleq \big(1+\boldsymbol r_{\mathrm B,n}^{T}\boldsymbol r_{\text{T},n}\big)^q/2^q\in[0,1]$
and
$b_n(\boldsymbol{l}_{\text{R}})\triangleq (d_{\mathrm B,n}^{-2}d_{\text{T},n}^{-2})$.
Since $\nabla v$ is Lipschitz on $\mathcal A$ with constant $L$ whenever
$\sup_{\boldsymbol{l}_{\text{R}}\in\mathcal A}\|\nabla^2 v(\boldsymbol{l}_{\text{R}})\|_2\le L$,
it suffices to upper bound $\|\nabla^2 c_n\|_2$ uniformly and sum over $n$.
Over $\mathcal A$, we have $d_{\mathrm B,n},d_{\text{T},n}\ge d_{\min}$ and $\|\boldsymbol r_{\mathrm B,n}\|=\|\boldsymbol r_{\text{T},n}\|=1$.
Moreover, the Jacobian of a unit direction vector satisfies
$\|\nabla_{\boldsymbol{l}_{\text{R}}}\boldsymbol r_{\mathrm X,n}\|_2\le 1/d_{\mathrm X,n}\le 1/d_{\min}$,
which yields $\|\nabla(\boldsymbol r_{\mathrm B,n}^{T}\boldsymbol r_{\text{T},n})\|_2\le 2/d_{\min}$
via (29).
Using standard derivative bounds for $d^{-2}_{\text{X},n}$, one obtains
$\|\nabla b_n\|_2\le 4/d_{\min}^5$ and $\|\nabla^2 b_n\|_2\le 20/d_{\min}^6$.
In addition, since $a_n(\cdot)$ is a $q$-th power of a bounded inner product, its first and second derivatives can be bounded as
$\|\nabla a_n\|_2\le q/d_{\min}$ and $\|\nabla^2 a_n\|_2\le (4q^2+14q)/d_{\min}^2$ by repeatedly applying the chain rule and the bounds above.
Finally, applying the product-rule decomposition
$\nabla^2 c_n=(\nabla^2 a_n)b_n+(\nabla a_n)(\nabla b_n)^T+(\nabla b_n)(\nabla a_n)^T+a_n\nabla^2 b_n$
and using $b_n\le 1/d_{\min}^4$ and $a_n\le 1$ yields the uniform bound
$\|\nabla^2 c_n\|_2\le (4q^2+22q+20)/d_{\min}^6$.
Summing over $n$ and multiplying by $\sqrt{\chi_0}$ proves \eqref{eq:L_global}.
\end{IEEEproof}
Now, maximizing \eqref{eq:mm_minorizer} over $\mathcal A$ yields a closed-form update
\begin{align}
\boldsymbol{l}_{\text{R}}^{(t+1)}
=
\Pi_{\mathcal A}\!\big(
\boldsymbol{l}_{\text{R}}^{(t)}
+L^{-1}\nabla_{\boldsymbol{l}_{\text{R}}}v(\boldsymbol{l}_{\text{R}}^{(t)})
\big),
\label{eq:mm_update}
\end{align}
where $\Pi_{\mathcal A}(\cdot)$ denotes the component-wise clipping onto $\mathcal A$.
This update guarantees monotonic improvement of the objective
$v(\boldsymbol{l}_{\text{R}}^{(t+1)})\ge v(\boldsymbol{l}_{\text{R}}^{(t)})$,
and equivalently $\gamma(\boldsymbol{l}_{\text{R}}^{(t+1)})\ge \gamma(\boldsymbol{l}_{\text{R}}^{(t)})$.
The complete three-stage optimization framework is summarized in Algorithm~\ref{alg:AO}.
In Stage~1, each MM iteration evaluates $\{d_{\mathrm B,n},d_{\text{T},n},\boldsymbol r_{\mathrm B,n},\boldsymbol r_{\text{T},n}\}_{n\in\mathcal N}$ and the gradient summation, leading to $\mathcal O(N)$ computational complexity per iteration.

\begin{algorithm}[t]
\caption{Overall Algorithm for (P1)}\label{alg:AO}
\begin{algorithmic}[1]
\STATE \textbf{Input:} $\boldsymbol{l}_{\text{B}}$, $\boldsymbol{l}_{\text{T}}$, $\{\boldsymbol{\bar l}_n\}_{n\in\mathcal N}$, $\mathcal A$, $q$, $\chi_0$, tolerance $\epsilon$.
\STATE \textbf{Stage 1:} Initialize $\boldsymbol{l}_{\text{R}}^{(0)}\in\mathcal A$ and set $t=0$.
\STATE \hspace{15pt} Compute $d_{\min}$ by \eqref{eq:dmin_def} and set $L$ by \eqref{eq:L_global}.
\STATE \hspace{15pt} \textbf{repeat}
\STATE \hspace{30pt} Update $\boldsymbol{l}_{\text{R}}^{(t+1)}$ by (34), and $t\leftarrow t+1$.
\STATE \hspace{15pt} \textbf{until} {$\|\boldsymbol{l}_{\text{R}}^{(t)}-\boldsymbol{l}_{\text{R}}^{(t-1)}\|\le \epsilon$}.
\STATE \textbf{Stage 2:} Given the optimal location $\boldsymbol{l}_{\text{R}}^\star=\boldsymbol{l}_{\text{R}}^{(t)}$, compute the optimal IRS phase shifts $\boldsymbol\theta^\star$ by (11) for phase alignment.
\STATE \textbf{Stage 3:} Given $\boldsymbol{l}_{\text{R}}^\star$, compute optimal rotations $\boldsymbol F^\star$ by (14).
\STATE \textbf{Output:} $\boldsymbol\theta^\star$, $\boldsymbol F^\star$, $\boldsymbol{l}_{\text{R}}^{\star}$.
\end{algorithmic}
\end{algorithm}

\vspace{-7pt}
\section{Numerical Results}
\vspace{-2pt}

\setlength{\abovecaptionskip}{-6pt}

\begin{figure}[t] 
\centering
\includegraphics[width=0.34\textwidth]{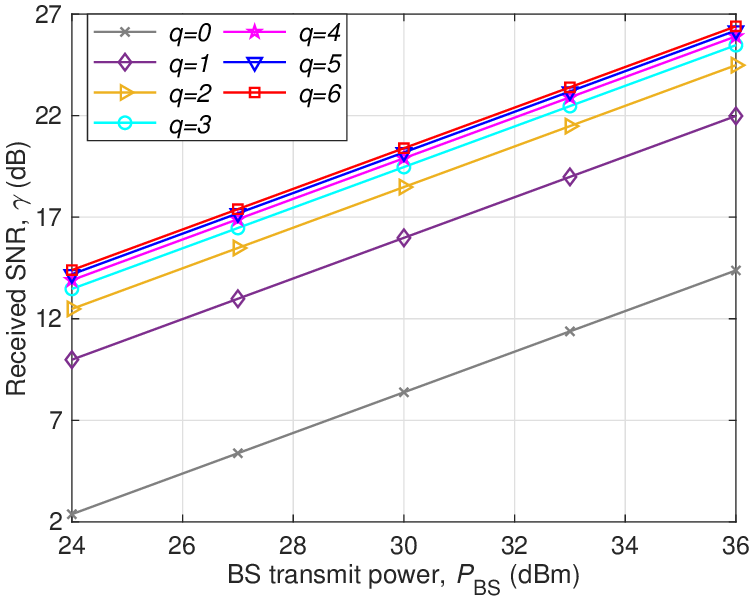}
\caption{SNR $\gamma$ versus BS transmit power with different directivity $q$.} 
\vspace{-21pt}
\end{figure}

\begin{figure*}[t] 
\centering
\includegraphics[width=0.85\textwidth]{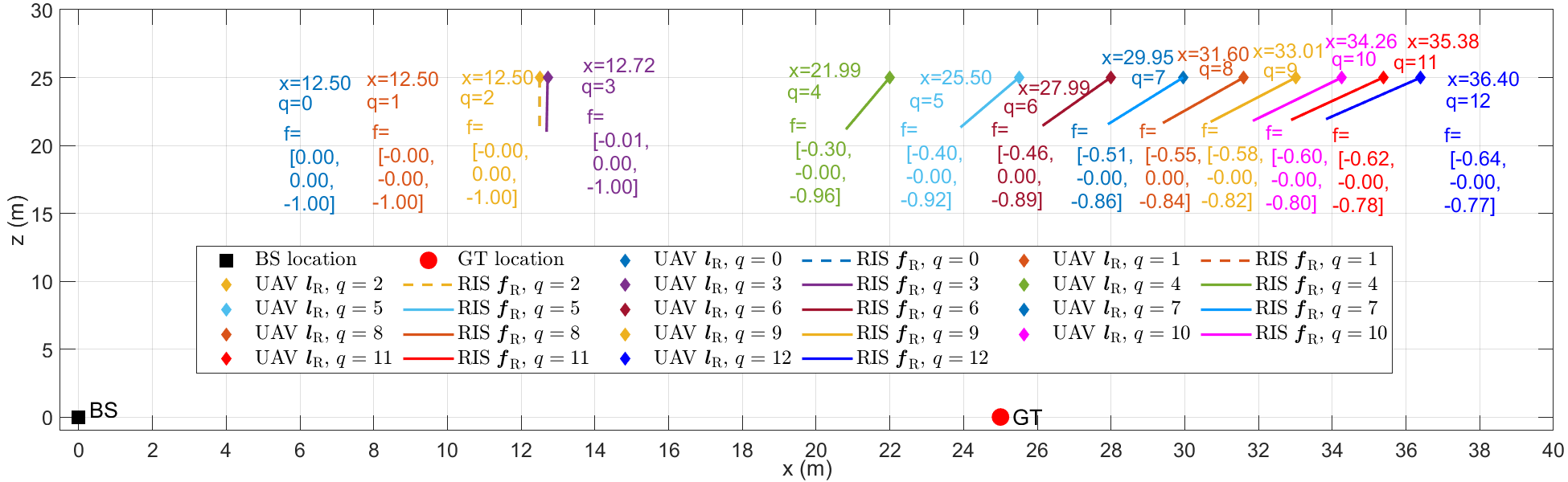}
\caption{A case study on the optimal location $\boldsymbol{l}_{\text{R}}^\star$ and rotation $\boldsymbol{f}_{\text{R}}^\star$ for maximizing SNR $\gamma$ with different directivity $q$.} 
\vspace{-18pt}
\end{figure*}

\begin{figure}[t]
\centering
\includegraphics[width=0.34\textwidth]{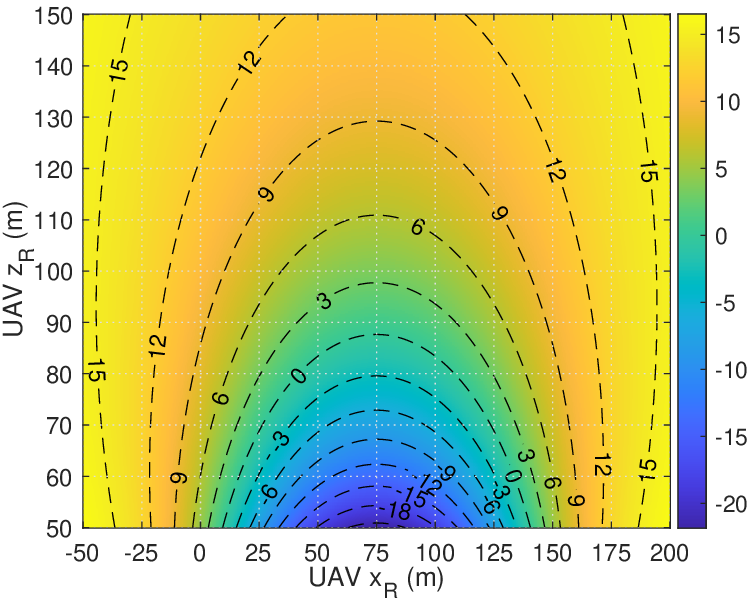}
\caption{Normalized SNR gain of $q=4$ versus $q=0$ over airspace.} 
\vspace{-18pt}
\end{figure}

Unless otherwise specified, the BS and GT are located on the ground with $\boldsymbol{l}_{\text{B}}=[0,0,0]^T$ and $\boldsymbol{l}_{\text{T}}=[x_{\text{T}},0,0]^T$. The certer location of UAV-mounted IRS $\boldsymbol{l}_{\text{R}}$ is constrained in a low-altitude airspace $\mathcal A$ with $x_{\text{R}}\in[-25,75]$~m, $z_{\text{R}}\in[25,50]$~m, and $y_{\text{R}}\in[-10,10]$~m. Fig. 2 plots the optimized received SNR $\gamma$ versus the BS transmit power $P_{\text{B}}$ for different directivity factors $q$ with a $20\times 20$ IRS and $x_{\text{T}}=25$~m. The clear monotonic improvement as $q$ increases confirms that directional radiation patterns of IRS elements can be exploited to strengthen the cascaded BS-UAV-GT channel in low-altitude communication. At $P_{\text{B}}=30$~dBm, the directional design with $q=6$ achieves about $12$~dB gain compared with the isotropic case $q=0$. Moreover, the performance gap between moderate-to-large $q$ values becomes small, suggesting a diminishing-return behavior, as the main-lobe is sufficiently narrow to concentrate energy along the two desired link directions.

To reveal the underlying mechanism behind the SNR gains, Fig. 3 shows a case study of the optimized UAV location $\boldsymbol{l}_{\text{R}}$ (all located onto the $xOz$ plane) and the corresponding rotated boresight of IRS elements, where the rotation pointing vector is computed by averaging $\{\boldsymbol f_n^\star\}$ over all elements, due to the numerically similar value between different elements. Notably, two important observations can be made. Firstly, when $q=0$, the element gain becomes independent of the pointing direction, and only the path-loss terms dominate. Accordingly, the optimized location is close to the midpoint between the BS and GT, i.e., $x_{\text{R}}=12.5$ for $x_{\text{T}}=25$, and tends to approach the lowest feasible altitude with $z_{\text{R}}= 25$, which jointly minimizes the product distance penalties in the cascaded link. This phenomenon aligns with the deployment insight in [5]. Secondly, directional IRS deliberately trades extra path loss for better angular alignment. Specifically, as $q$ increases, the optimal placement shifts toward the GT side with larger $x_{\text{R}}$ (also has the same performance symmetry to $(x_{\text{B}}+x_{\text{T}})/2=25$ toward the BS side), while the optimal boresight tilts toward the BS, as explicitly shown by the labeled vectors in Fig. 3. This behavior is consistent with Proposition~1, where $\boldsymbol f_n^\star$ points along the angle bisector of the BS and GT directions, i.e., it balances and improves the two cosine gains simultaneously. Geometrically, moving the UAV-mounted IRS toward one terminal reduces the angular separation between the two link directions, as seen from the IRS, thus increasing the factor $(1+\boldsymbol r_{\mathrm B,n}^T\boldsymbol r_{\text{T},n})^{2q}$ embedded in the large-scale gain after substituting $\boldsymbol f_n^\star$. However, this comes at the cost of increased BS-IRS and IRS-GT distance (and hence larger productive path loss). Therefore, Fig. 3 unveils a key deployment insight in low-altitude UAV-IRS communication: directional elements tend to favor a deployment closer to one node and even exceed the area between BS and GT, to harvest stronger directional gain, even if the resulting BS-side path loss becomes larger than that of the isotropic optimum, and the net effect still improves the received SNR. This fundamental tradeoff between directional gain and path loss does not appear in conventional isotropic IRS systems and is intrinsic to UAV-mounted rotatable IRS with directive patterns.

Finally, Fig. 4 characterizes where directional IRS is more or less beneficial by plotting the normalized SNR gain, defined as $10\log_{10}\!\left(\gamma|_{q=4}\big/\gamma|_{q=0}\right)$, over a wide airspace region for a longer BS-GT separation $x_{\text{T}}=150$~m and a larger $N=60\times 60$ IRS with $y_{\text{R}}=0$. At each given airspace location, the directional case $q=4$ uses the closed-form $\{\boldsymbol f_n^\star\}$, while $q=0$ corresponds to the isotropic benchmark. Two regimes are clearly observed. The first one is the directional-favorable regime: when the UAV-IRS location makes the BS and GT directions as seen from the IRS moderately aligned (e.g., the UAV is sufficiently high or located away from the midpoint of the BS-GT segment), the directional design yields a positive normalized gain, indicating that directive gain dominates the additional path loss in these geometries. The second is the isotropic-favorable regime: when the UAV-IRS is in a geometry in which the BS and GT directions subtend a large angle at the IRS (typically near the middle of a long BS-GT segment at very low altitude), the directional advantage diminishes and can even become marginal compared with the isotropic case.
This happens because a highly directive element cannot simultaneously maintain large cosine projections onto two widely separated directions, making the product-form directional gain less effective. Overall, Fig. 4 provides a practical airspace-level guideline for low-altitude deployment: directional IRS is more suitable when the UAV can be positioned such that the two hop directions are not excessively separated in angle; otherwise, an isotropic or weakly directive IRS may be preferable.

\vspace{-3pt}
\section{Conclusion}
\vspace{-3pt}
This paper investigated low-altitude reflection with a UAV-mounted rotatable IRS under a practical hemispherical directive element pattern. By jointly designing the location, rotation vectors, and phase shifts, we derived a closed-form phase alignment and uncovered a geometric rotation rule: each element's optimal boresight aligns with the internal angular bisector between the BS-IRS and GT-IRS directions. With these closed-form solutions, the remaining placement design was efficiently addressed using an algorithm based on MM and the projected gradient method to maximize SNR subject to box constraints. Numerical results verified substantial SNR gains from directive elements and highlighted a fundamental trade-off between directional gain and path loss, yielding new insights into low-altitude deployment of UAV-mounted IRS. 

\ifCLASSOPTIONcaptionsoff
  \newpage
\fi

\end{document}